\documentclass[graybox]{svmult}

\usepackage{mathptmx}       
\usepackage{helvet}         
\usepackage{courier}        
\usepackage{type1cm}        
\usepackage{makeidx}         
\usepackage{graphicx}        
\usepackage{multicol}        
\usepackage[bottom]{footmisc}

\usepackage{amssymb}

\makeindex             

\begin{document}

\title*{The effect of the environment on the gas kinematics and morphologies of distant galaxies}
\titlerunning{Effect of environment on the gas and stars of distant galaxies} 
\author{Yara L. Jaff\'e}
\institute{Yara L. Jaff\'e \at School of Physics \& Astronomy, The University of Nottingham, University Park, Nottingham NG7 2RD, UK \email{ppxyj@nottingham.ac.uk}}
%
%
\maketitle

\vskip-1.2truein

\abstract{ With the aim of understanding which physical processes are primarily responsible for the transformation of spiral galaxies into S0s in clusters, we study the gas kinematics, morphological disturbances, and the Tully-Fisher relation (TFR) of distant galaxies in various environments. We use the ESO Distant Cluster Survey (EDisCS) dataset, that spans a broad range of cluster and galaxy properties at $0.3 < z < 0.8$. Our results indicate that the physical mechanism acting on cluster galaxies (with $M_B \leqslant -20$mag) must be strong enough to significantly disturb the gas in cluster galaxies, but at the same time, mild enough to leave the stellar structure unaffected.\newline}


Over the past decades, much observational evidence  have suggested  the transformation of star-forming spirals into passive S0s by the influence of the cluster environment. However, the physical processes responsible for this transformation remain unclear. A number of plausible mechanisms have been proposed. These include, ram-pressure stripping \cite{GunnGott1972}, mergers \cite{Bekki1998}, galaxy harassment \cite{Moore1999} and tidal stripping \cite{Larson1980,Balogh2000}. Each one of these mechanisms is expected to be effective in different regions of the cluster environment and affect galaxy properties in different ways. For example, a potential key difference between the ram pressure stripping and the merger or tidal stripping scenario is that the former is likely to enhance star formation across the disk \cite{BekkiCouch2003}, while mergers or tides could cause a starburst that is probably centrally concentrated \cite{MihosHernquist1994}.
To attack this problem, we study the effect of environment on the gas kinematics of distant galaxies as well as the morphological disturbances. We focus on a luminosity-limited ($M_B \leqslant -20$mag) sub-sample of emission-line galaxies from the EDisCS dataset (see \cite{white,Halliday2004,MJ2008,Desai2007}) at $0.3<z<0.8$. Our sample consists of 224 spectroscopically confirmed cluster and field galaxies.  From the emission-lines in the spectra, we measure rotation velocities and identify galaxies with significant kinematical disturbance. For most of the sample we also have HST imaging from which we are able to quantify morphological disturbances. 

\begin{figure*}[t]
%
\includegraphics[width=0.49\textwidth]{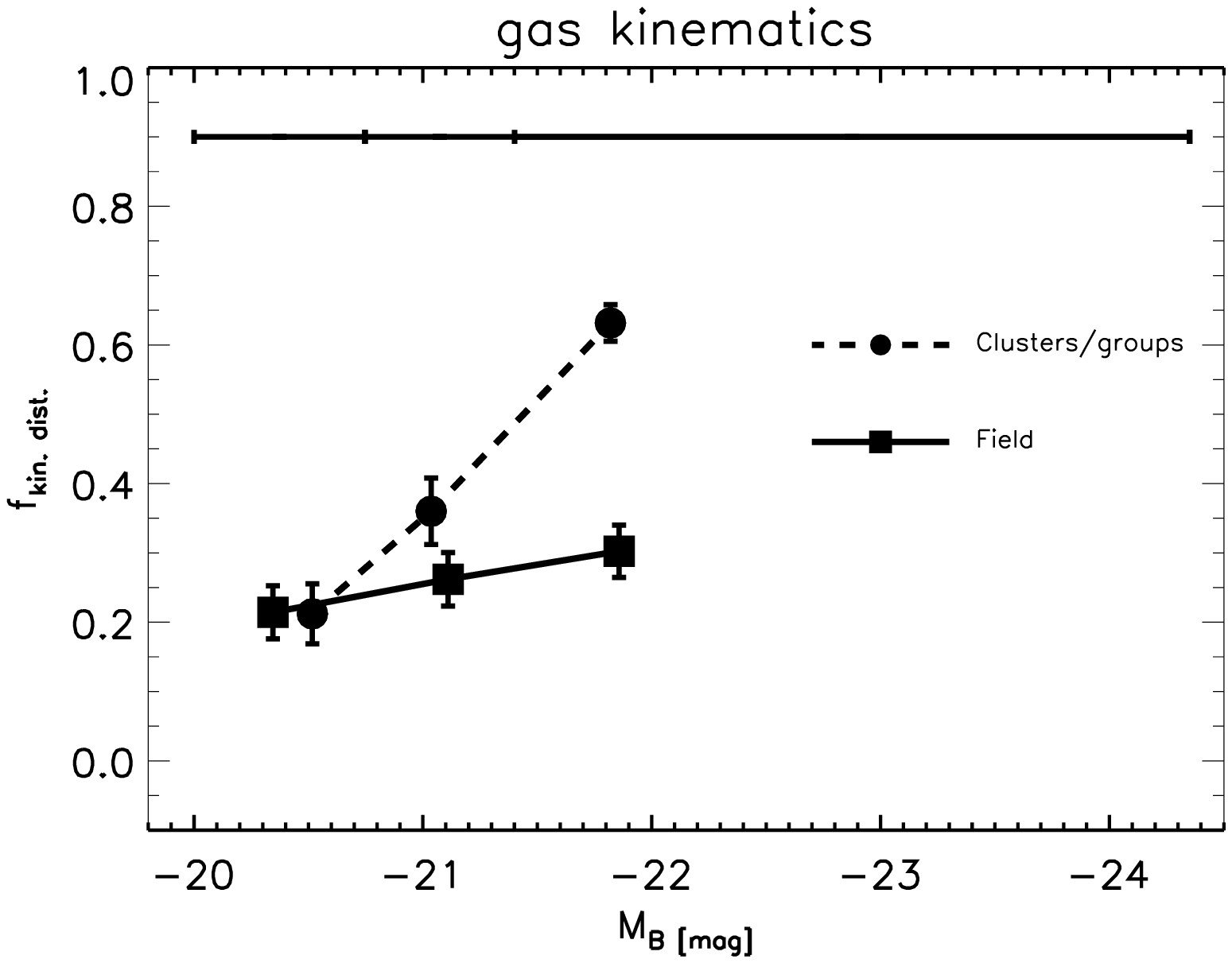}
\includegraphics[width=0.49\textwidth]{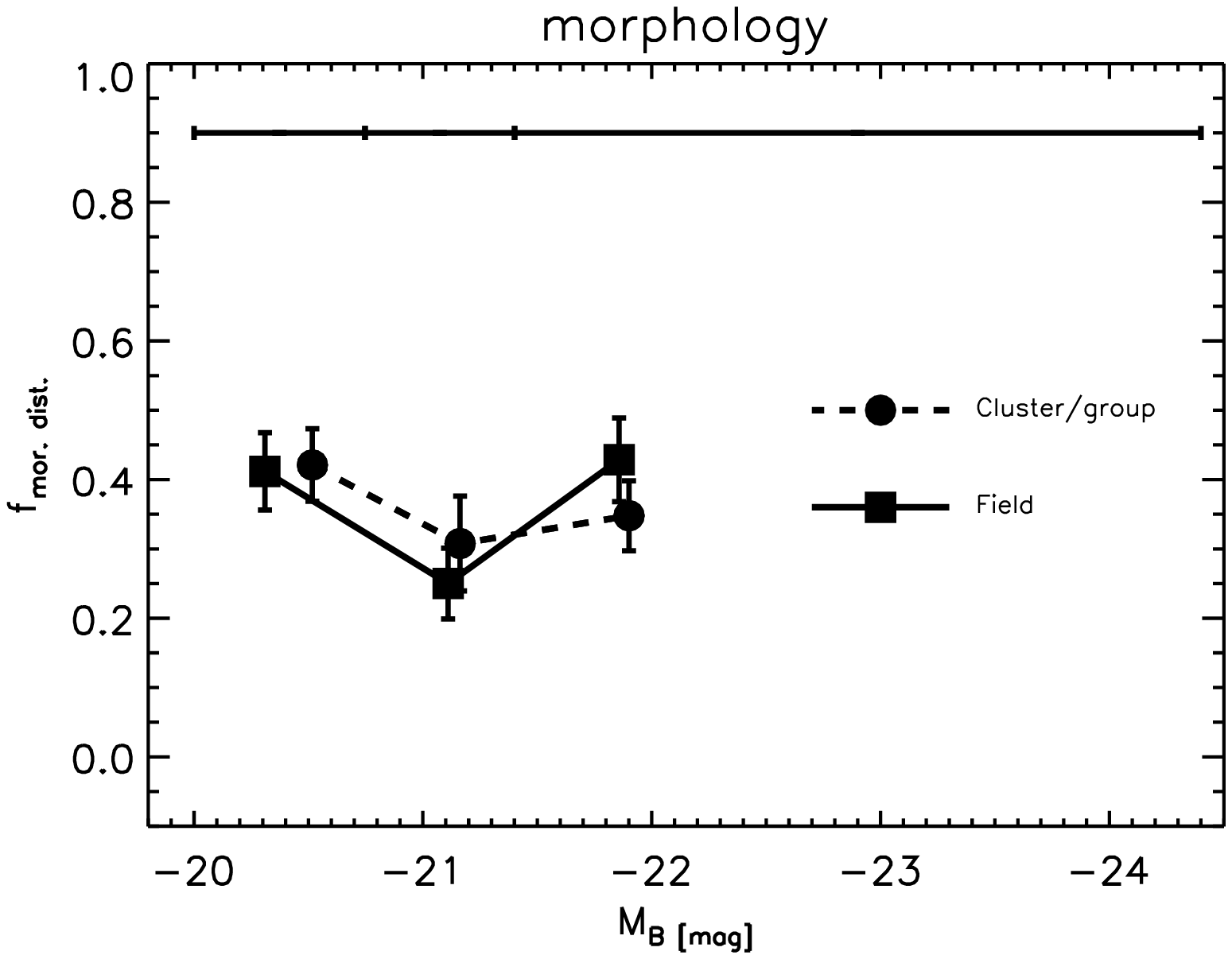} \caption{ Fraction of kinematically disturbed (left) and morphologically disturbed (right) galaxies as a function of $M_B$ for different environments (labeled). $M_B$ bins are shown at the top of each panel.}
\label{fig:Jaffe1}       
\end{figure*}

Our main result is shown in Figure~\ref{fig:Jaffe1}, where the fraction of kinematically disturbed  galaxies (left),  and the fraction of galaxies with disturbed morphologies (right) are plotted as a function of rest frame $B$-band magnitude and environment. It is clear that there are significantly more kinematically disturbed galaxies in cluster environment than in the field, and that the morphological disturbance does not seem to be affected by environment. We interpret the rise of the kinematical disturbance with brighter $M_B$ as a result of smaller cluster galaxies being already stripped off their gas and thus not present in our (emission-line) sample.

We also compare the TFR of cluster and field galaxies (in matched samples of z and $M_B$) of those galaxies with no signs of kinematical disturbance (i.e. reliable rotation velocities). At face value, we conclude that there are no environmental effects on the TFR, although we are currently investigating the star formation properties of the kinematically disturbed  galaxies to test if they could enhance $M_B$ in the TFR.

Finally, we found that although the vast majority of our emission-line galaxies are spirals and a few irregulars, there are also early-type galaxies, some of which have an extended gas disk in their emission. We are currently investigating the origin of the gas in these objects and their link to lower redshift analogs (e.g. \cite{emsellem07}).

The results discussed above, and other related results, are discussed in more detail in Jaff\'e et al.~2010a, 2010b, in prep.)
\begin{acknowledgement}
Based on observations collected at the European Southern Observatory, Chile, as part of programme 073.A-0216. The author is greatly indebted to the European Southern Observatory, the Royal Astronomical Society and the astrophysics group of Exeter University.
\end{acknowledgement}
%


\end{document}